\documentclass[11pt]{article}
\usepackage{amssymb}  

\usepackage{amsmath, amsthm, amssymb}
\usepackage{graphicx}
\usepackage{multirow}
\usepackage{xr}

\usepackage{cite}
\makeatletter
\renewcommand{\@biblabel}[1]{\quad#1.}
\makeatother

\usepackage[labelfont=bf,labelsep=period,justification=raggedright]{caption}

\theoremstyle{definition}

\theoremstyle{remark}

\newcommand{\EV}{{\bf E}} 
\newcommand{\bfy}{{\bf y}} 
\newcommand{\bfI}{{\bf I}} 
\newcommand{\bfr}{{\bf r}} 
\newcommand{\bfK}{{\bf K}} 
\newcommand{\bfJ}{{\bf J}} 
\newcommand{\bfC}{{\bf C}}
\newcommand{\bfA}{{\bf A}}
\newcommand{\bfW}{{\bf W}}
\newcommand{\bfX}{{\bf X}}
\newcommand{\bfY}{{\bf Y}}
\newcommand{\bfH}{{\bf H}}
\newcommand{\bfone}{{\bf 1}}
\newcommand{\bfeta}{{\boldsymbol\eta}}
\newcommand{\bfrho}{{\boldsymbol\rho}}

\newcommand{\At}{\tilde{A}}

\newcommand{\Kt}{\tilde{K}}
\newcommand{\Jt}{\tilde{J}}
\newcommand{\Ct}{\tilde{C}}
\newcommand{\mut}{\tilde{\mu}}

\newcommand{\bfyt}{{\bf \tilde{y}}} 
\newcommand{\bfKt}{{\bf \tilde{K}}} 
\newcommand{\bfId}{{\bf I}} 
\newcommand{\bfCt}{{\bf \tilde{C}}}
\newcommand{\bfAt}{{\bf \tilde{A}}}
\newcommand{\bfJt}{{\bf \tilde{J}}}

\newcommand{\taur}{\tau_{\rm ref}} 
\newcommand{\vth}{v_{\rm th}} 
\newcommand{\vr}{v_{\rm r}} 

\newcommand{\cov}{\text{cov}}
\newcommand{\var}{\text{var}}

\newcommand{\beq}{\begin{equation}}
\newcommand{\eeq}{\end{equation}}
\newcommand{\beqr}{\begin{eqnarray}}
\newcommand{\eeqr}{\end{eqnarray}}
\newcommand{\beqrn}{\begin{eqnarray*}}
\newcommand{\eeqrn}{\end{eqnarray*}}
\newcommand{\beqn}{\begin{equation*}}
\newcommand{\eeqn}{\end{equation*}}
\newcommand{\bei}{\begin{itemize}}
\newcommand{\beii}{\begin{itemize} \item}
\newcommand{\eei}{\end{itemize}}
\newcommand{\ben}{\begin{enumerate}}
\newcommand{\een}{\end{enumerate}}
\newcommand{\bes}{\begin{small}}
\newcommand{\ees}{\end{small}}
\newcommand{\bec}{\begin{center}}
\newcommand{\eec}{\end{center}}

\newcommand{\lsim}{\mathrel{\hbox{\rlap{\lower.55ex \hbox{$\sim$}} \kern-.3em \raise.4ex \hbox{$<$}}}}
\newcommand{\gsim}{\mathrel{\hbox{\rlap{\lower.55ex \hbox{$\sim$}} \kern-.3em \raise.4ex \hbox{$>$}}}}

\usepackage[usenames,dvipsnames]{color}

\newcommand{\EVb}[1]{\EV\hspace{-0.03in}\left\{#1\right\}} 
\newcommand{\EVs}[1]{\EV\hspace{-0.03in}\left[#1\right]} 

\title{Impact of network structure and cellular response on spike time correlations}

\author{
James Trousdale\thanks{Department of Mathematics, University of Houston, Houston, TX 77204-3008, USA}, 
Yu Hu\thanks{Department of Applied Mathematics, Program in Neurobiology and Behavior, University of Washington, Seattle, WA 98195}, Eric Shea-Brown$^\dagger$ and Kre\v{s}imir Josi\'{c}$^*$\thanks{Department of Biology and Biochemistry, University of Houston, Houston, TX 77204-5001}
}

\begin{document}

\begin{flushleft}
{\Large
\textbf{Impact of network structure and cellular response on spike time correlations}
}
\\
James Trousdale$^1$,
Yu Hu$^3$,
Eric Shea-Brown$^3$,
Kre\v{s}imir Josi\'{c}$^{1,2}$
\\
{\bf1} Department of Mathematics, University of Houston, Houston, TX, USA
\\
{\bf2} Department of Biology and Biochemistry, University of Houston, Houston, TX, USA
\\
{\bf3} Department of Applied Mathematics, Program in Neurobiology and Behavior, University of Washington, Seattle, WA, USA\\
\end{flushleft}

\renewcommand{\arraystretch}{2}

\section*{Abstract}

Novel experimental techniques reveal the simultaneous activity of larger and larger numbers of neurons.  As a result there is increasing interest in the structure of cooperative -- or {\it correlated} -- activity in neural populations, and in the possible impact of such correlations on the neural code.   A fundamental theoretical challenge is to understand how the architecture of network connectivity along with the dynamical properties of single cells  shape the magnitude and timescale of correlations.  We provide a general approach to this problem by extending prior techniques based on {\it linear response theory}.  We consider networks of general integrate-and-fire cells with arbitrary architecture, and provide explicit expressions for the approximate cross-correlation between constituent cells.  These correlations depend strongly on the operating point (input mean and variance) of the neurons, even when connectivity is fixed.  Moreover,  the approximations admit an expansion in powers of the matrices that describe the network architecture.  This expansion can be readily interpreted in terms of paths between different cells.  We apply our results to large excitatory-inhibitory networks, and demonstrate first how precise {\it balance}  --- or lack thereof --- between the strengths and timescales of excitatory and inhibitory synapses is reflected in the overall correlation structure of the network. We then derive explicit expressions for the average correlation structure in randomly connected networks. These expressions help to identify the important factors that shape coordinated neural activity in such networks.

\section*{Author summary}

Is neural activity more than the sum of its individual parts?  What is the impact of cooperative, or {\it correlated}, spiking among multiple cells?  We can start addressing these questions, as rapid advances in experimental techniques allow simultaneous recordings from ever-increasing populations.  However, we still lack a general understanding of the origin and consequences of the joint activity that is revealed.  The challenge is compounded by the fact that both the intrinsic dynamics of single cells and the correlations among then vary depending on the overall state of the network.  Here, we 
develop a toolbox that addresses this issue.  Specifically, we show how {\it linear response theory} allows for the 
expression of correlations explicitly in terms of the underlying network connectivity and known single-cell
properties --- and that the predictions of this theory accurately match simulations of a touchstone, nonlinear model in computational neuroscience, the general integrate-and-fire cell.  Thus, our theory should help unlock the relationship between network architecture, single-cell dynamics, and correlated activity in diverse neural circuits.

\section*{Introduction}

New multielectrode and imaging techniques are revealing the simultaneous activity of neural ensembles and, in some cases, entire neural populations~\cite{Cohen:2011,Ohki:2005,Field:2010,Jia:2010}.  This has thrust upon the computational biology community the challenge of characterizing a potentially complex set of interactions --- or {\it correlations} --- among pairs and groups of neurons.  

Beyond important and rich challenges for statistical modeling~\cite{BrownKM04}, the emerging data promises new perspectives on the neural encoding of information~\cite{Averbeck:2006}.  The structure of correlations   in the activity of neuronal populations is of central importance in understanding the neural code~\cite{Shadlen:1998,PanzeriSTR99,Zoh+94,abbott99,sompolinsky01,Pan+01,Beck:2011}.  However, theoretical~\cite{JosicSDR08,Zoh+94,abbott99,sompolinsky01,Sch+03,Lat+05}, and empirical studies~\cite{Nir+01,Cohen:2009,RHZE03} do not provide a consistent set of general principles 
about the impact of correlated activity. This is largely because the presence of correlations can either strongly increase or decrease the fidelity of encoded information  depending on both the structure of correlations across a population and how their impact is assessed.

A basic mechanistic question underlies the investigation of the role of collective activity in coding and signal transmission: How do single-cell dynamics, connection architecture, and synaptic dynamics combine to determine patterns of network activity?  Systematic answers to this question would allow us to predict how empirical data from one class of stimuli will generalize to other stimulus classes and recording sites.    Moreover, a mechanistic understanding of the origin of correlations, and knowledge of the patterns we can expect to see under different assumptions about the underlying networks,   will help resolve recent controversies about the strength and pattern of correlations in mammalian cortex ~\cite{Renartetal10,Cohen:2011,Ecker:2010,Dragoi:2011}.  Finally, understanding the origin of correlations will inform the more ambitious aim of inferring properties of network architecture from  observed patterns of activity~\cite{Paninski:2010,Nykamp:2007, Ostojic:2009}.

Here, we examine the link between network properties and correlated activity.  We develop a theoretical framework that accurately predicts the structure of correlated spiking that emerges in a widely used model --- recurrent networks of general integrate and fire cells.  The theory naturally captures the role of single cell and synaptic dynamics  in shaping the magnitude and timescale of spiking correlations.  We focus on the exponential integrate and fire model, which has been shown to capture membrane and spike responses of cortical neurons~\cite{FourcaudTrocme:2003}; however, the general approach we take can be applied to a much broader class of neurons, a point we return to in the Discussion.

Our approach is based on an extension of linear response theory to  networks~\cite{Lindner:2005,Ostojic:2009}.  We start with a linear approximation of a neuron's response to an input.  
This approximation can be obtained explicitly for many neuron models~\cite{Brunel:1999,Lindner:2001,Richardson:2007}, and is directly related to
the spike triggered average~\cite{Gabbiani:1998}.  The correlation structure of the network is then estimated using an iterative approach.   As in prior work~\cite{Pernice:2011,Rangan:2009-1,Rangan:2009-2}, the resulting expressions admit an expansion in terms of paths through the network.

We apply this theory to networks with precisely balanced inhibition and excitation in the inputs
to individual cells.  In this state  individual cells receive a combination of excitatory and inhibitory inputs with mean values that largely cancel.  We show that, when timescales and strengths of excitatory and inhibitory connections are matched, only local interactions
between cells contribute to correlations.    Moreover, our theory allows us to explain how correlations are altered when precise tuning balance is broken.  In particular, we show how strengthening inhibition may synchronize the spiking activity 
in the network. Finally, we derive results which allow us to gain an intuitive understanding of the factors shaping average correlation structure in randomly connected networks of neurons.

\section*{Network model and linear response theory}

Our goal is to understand how the architecture of a network shapes the statistics of
its activity.   We show how correlations between spike trains of cells can  be approximated using 
response characteristics of individual cells along with information about synaptic dynamics, and the structure of the network.  We start by briefly reviewing linear response theory of neuronal responses~\cite{Risken:1996,Lindner:2001,Brunel:2001}, and then use it to approximate the correlation structure of a network.

\subsection*{Network model}

 To illustrate the results we
consider a network of $N$ nonlinear integrate-and-fire (IF) neurons with membrane
potentials modeled by
\begin{equation} \label{E:main}
\tau_i\dot{v_i} = -(v_i-E_{L,i}) + \psi(v_i) + E_i + \sqrt{\sigma_i^2\tau_i}\xi_i(t) +   f_i(t) + \eta_i(t).
\end{equation}
Here 
$E_i$ represents the mean synaptic input current from parts of the system
not explicitly modeled.  A spike-generating current $\psi(v_i)$ may be included to emulate the rapid onset of action potentials. Unless otherwise specified, we utilize the exponential IF model (EIF), so that
$\psi(v) \equiv \Delta_T \exp[(v-v_T)/\Delta_T]$~\cite{FourcaudTrocme:2003}.  Cells are subject to internally induced fluctuations due to channel noise~\cite{White:2000}, and externally induced fluctuations due to inputs not explicitly modelled~\cite{Renart:2004}.   We
model both by independent,
Gaussian, white noise processes, $\sqrt{\sigma_i^2\tau_i}\xi_i(t)$~\cite{Burkitt:2006}.   An external signal to  cell $i$ is 
represented  by $\eta_i(t)$.  

Upon reaching a threshold $\vth$, an action potential is generated, and the membrane potential is reset to
$\vr$, where it is held constant for an absolute refractory period $\taur$.
The output of cell $i$ is characterized by the times, $t_{i,k}$, at which its membrane
potential reaches threshold, resulting in an output spike train
$
y_i(t) = \sum_k \delta(t-t_{i,k}).
$
Synaptic interactions are modeled by delayed $\alpha$-functions
\begin{equation} \label{E:synaptic}
f_i(t) = \sum_j ( \bfJ_{ij} * y_j)(t),\qquad\text{where}\qquad \bfJ_{ij}(t) = \begin{cases} 
\bfW_{ij} \left(\frac{t-\tau_{D,j}}{\tau_{S,j}^2}\right)\exp\left[-\frac{t-\tau_{D,j}}{\tau_{S,j}}\right]&\qquad t \geq \tau_{D,j} \\ 
0 & \qquad t < \tau_{D,j}
\end{cases}.
\end{equation}
The $N \times N$ matrix $\bfJ$ contains the synaptic kernels, while the matrix $\bfW$ contains the synaptic weights, and hence defines the network architecture.  In particular,  $\bfW_{ij} =0$ represents the absence of a
synaptic connection from cell $j$ to cell $i$. 

Table~\ref{T:notation} provides an overview of all parameters and variables.

\subsection*{Measures of spike time correlation}

We quantify  dependencies between the  responses of cells in the network
using the spike train auto- and cross-correlation functions~\cite{Gabbiani:2010}. For a pair of spike trains, $y_i(t), y_j(t)$, the cross-correlation
function $\bfC_{ij}(\tau)$ is defined as
$$
\bfC_{ij}(\tau) = \cov\left(y_i(t+\tau),y_j(t)\right).
$$
The auto-correlation function $\bfC_{ii}(t)$ is  the cross-correlation between
a spike train and itself, and $\bfC(t)$ is the matrix of cross-correlation functions.  Denoting
 by $N_{y_i}(t_1,t_2) = \int_{t_1}^{t_2} y_i(s)ds$ the number of spikes over a time
window $[t_1,t_2]$, 
the spike count correlation, $\bfrho_{ij}(T)$, over windows
of length $\tau$ is defined as, 
$$
\bfrho_{ij}(T) = \frac{\cov\left(N_{y_i}(t,t+T),N_{y_j}(t,t+T)\right)}{\sqrt{\var\left(N_{y_i}(t,t+T)\right) \var\left(N_{y_j}(t,t+T)\right)}}.
$$
We assume stationarity of the spiking processes (that is, the network has reached a steady state) so that $\bfrho_{ij}(T)$ does not depend on $t$. We also use the total correlation coefficient $\bfrho_{ij}(\infty) =  \lim_{T\rightarrow \infty} \bfrho_{ij}(T)$ to characterize dependencies between the processes $y_i$ and $y_j$ over arbitrarily long timescales. 

The spike count covariance is related to the cross-correlation function by~\cite{Shadlen:1998,Bair:2001}
$$
\cov\left(N_{y_i}(t,t+\tau),N_{y_j}(t,t+\tau)\right) = \int_{-\tau}^\tau \bfC_{ij}(s)(\tau - |s|)ds.
$$

We can interpret the cross-correlation as the conditional probability that cell $j$ spikes at time $t+\tau$ given that cell $i$ spiked at time $t$.
The conditional firing rate,
$$
\bfH_{ij}(\tau) = \lim_{\Delta t \rightarrow 0} \frac{1}{\Delta t} \text{Pr}\left( N_{y_j}(t+\tau,t+\tau+\Delta t) > 0 \ | \ N_{y_i}(t,t+\Delta t) > 0\right),
$$
is the firing rate of cell $j$ conditioned on a spike in cell $i$ at  $\tau$ units of time in the past, and
$
\bfC_{ij}(\tau) = r_i(\bfH_{ij}(\tau) - r_j).
$

\subsection*{Linear response of individual cells}

Neuronal network models are typically described by a complex system of coupled nonlinear stochastic differential equations. Their behavior is therefore difficult to analyze directly.     We will use
linear response theory~\cite{Gabbiani:2010,Risken:1996,Lindner:2001,Brunel:2001} to approximate the cross-correlations between the outputs of neurons in a network.  We first review the 
linear approximation to the response of a single cell. We illustrate the approach using current-based IF neurons, and explain how it can be generalized to  other models in the Discussion. 

\begin{figure}
\centering
\includegraphics[scale = 1]{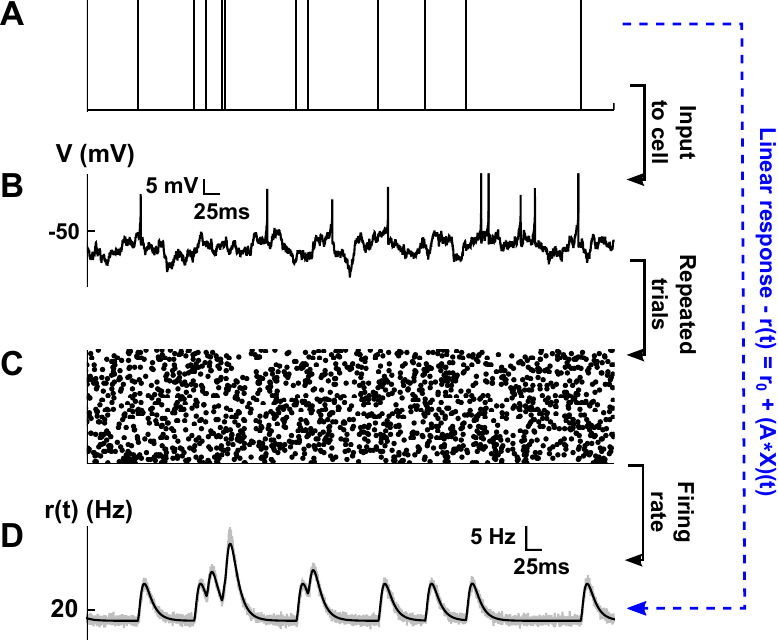}
\caption{{\bf Illustrating Eq.~\eqref{E:r_approx}. }
{\bf (A)} The input  to the post-synaptic cell is a fixed spike train which is convolved with a synaptic kernel. 
{\bf (B)} A sample voltage path for the post-synaptic cell receiving the input shown in A) in the presence of background noise.
{\bf (C)} Raster plot of 100 realizations of output spike trains of the post-synaptic cell.  
{\bf (D)} The output firing rate, $r(t)$, obtained by averaging over realizations of the output spike trains  in C). The rate obtained using Monte Carlo simulations (shaded in gray) matches  predictions of   linear response theory obtained using  Eq.~\eqref{E:r_approx} (black). 
}
\label{F:th_fig}
\end{figure}

The membrane potential of an IF neuron receiving input $\epsilon X(t)$, with vanishing temporal average, $\langle X(t) \rangle = 0$, evolves according to
\begin{equation}\label{E:v_theory}
\tau \dot{v} = -(v-E_L)+ \psi(v) + E + \sqrt{\sigma^2\tau} \xi(t) + \epsilon X(t).
\end{equation}
The time-dependent firing rate, $r(t)$, is determined by averaging the resulting spike train, $y(t) = \sum_j \delta(t-t_j),$  across different realizations of noise, $\xi(t),$ for fixed $X(t)$. 
Using linear response theory, we can approximate the firing rate by
\begin{equation}\label{E:r_approx}
r(t) = r_0 + (A*\epsilon X)(t),
\end{equation}
where $r_0$ is the (stationary) firing rate when $\epsilon = 0$. The linear response kernel, $A(t),$ characterizes  the firing rate response to first order in $\epsilon$. A rescaling of the function $A(t)$ gives the spike-triggered average of the cell, to first order in input strength, and is hence  equivalent to the optimal Weiner kernel in the presence of the signal $\xi(t)$.~\cite{Gabbiani:2010,Barreiro:2010}.  In Fig.~\ref{F:th_fig}, we compare the approximate firing rate obtained from Eq.~\eqref{E:r_approx} to that obtained numerically from Monte Carlo
simulations.

The linear response kernel $A(t)$ depends implicitly on model
parameters, but is independent of the input signal, $\epsilon X(t)$, when $\epsilon$ is small relative to the noise $\sqrt{\sigma^2 \tau}\xi(t)$.
In particular, $A(t)$ is sensitive 
to the value of the mean input current, $E$.  We emphasize that the presence of the background noise, $\xi$, in Eq.~\eqref{E:v_theory} is essential to the theory, as noise linearizes the transfer
function that maps input to output.

\subsection*{Linear response in recurrent networks}

The linear response kernel can be used to approximate the response of a cell
to an external input.  However, the situation is more complicated in a network where a neuron 
 can affect its own activity through recurrent connections.
To extend the linear response approximation to networks we follow the approach introduced by~\cite{Lindner:2005}. Instead of using the linear response kernel to approximate the firing rate of a cell, we use it to approximate a realization of its output
\begin{equation}\label{E:approx}
y(t) \approx y^1(t) = y^0(t) +(A*X)(t).
\end{equation}
Here $y^0(t)$ represents a realization of the spike train generated by an integrate-and-fire neuron obeying Eq.~\eqref{E:v_theory} with $X(t) = 0$.

The central assumption we make is that a cell acts approximately as a  linear filter of its inputs. Note that Eq.~\eqref{E:approx} defines a mixed point and continuous process, but
averaging $y(t)$ in Eq.~\eqref{E:approx} over realizations of $y^0$ gives  Eq.~\eqref{E:r_approx}.
Hence, Eq.~\eqref{E:approx} can be viewed as a natural generalization of Eq.~\eqref{E:r_approx} where the unperturbed output of the cell is represented as a point process, $y^0(t)$, instead of the firing rate, $r_0$.

We first use Eq.~\eqref{E:approx} to describe spontaneously evolving networks  where $\eta_i(t) = 0$.    Equation~\eqref{E:main} can
then be rewritten as
\begin{equation}\label{E:main2}
\tau_i\dot{v_i} = -(v_i-E_{L,i}) + \psi(v_i) + E_i' + \sqrt{\sigma_i^2\tau_i} \xi_i(t) + ( f_i(t) - \EVs{f_i}),
\end{equation}
where $E_i' = E_i +\EVs{f_i}$  and $\EVs{\mathbf{\cdot}}$ represents the temporal average.

As a first approximation of the spiking output of cells in the coupled network, we start with
realizations of spike trains, $y_i^0$,  generated by  IF neurons
obeying Eq.~\eqref{E:main2} with $f_i(t) = \EVs{ f_i}$. This is equivalent to considering
neurons isolated from the network, with adjusted DC inputs (due to mean network interactions).   Following the approximation given by Eq.~\eqref{E:approx},
we use a \emph{frozen} realization of all $y_i^0$ to find a correction to the output of each
cell, with $X(t)$ set to the mean-adjusted synaptic input, 
$$X(t) = f_i(t) - \EVs{ f_i}.$$  
As noted previously, the linear response kernel is sensitive to changes in the mean
input current.  It is therefore important to include the average synaptic input $\EVs{f_i}$ in the
definition of the effective mean input,  $E_i'$.

The input from cell $j$ to cell $i$ is filtered by the synaptic kernel $\bfJ_{ij}(t)$.  The linear response
of cell $i$ to a spike in cell $j$ is therefore captured by the interaction kernel $\bfK_{ij}$ defined by
$$
\bfK_{ij}(t) \equiv (A_i*\bfJ_{ij}) (t).
$$
The output of cell $i$ in response to mean-adjusted input, $y_j^0(t) - r_j$, from cell $j$
can be approximated to first order in input strength using the
linear response correction
\begin{equation}\label{eq:first_it_def}
y_i^1(t) = y_i^0(t) + \sum_j (\bfK_{ij}*[y_j^0 - r_j])(t).
\end{equation}
We explain how to approximate the stationary rates, $r_j$, in the Methods section.

%
%

The cross-correlation between the processes $y_i^1(t)$ in Eq.~\eqref{eq:first_it_def} gives
a first approximation to the cross-correlation function  between the cells (See Methods),
\begin{equation} \label{E:first}
\begin{split}
\bfC_{ij}(\tau)  \approx \bfC^1_{ij}(\tau) &= \EVs{ (y_i^1(t+\tau)-r_i)  (y_j^1(t)-r_j)}   \\
& = \delta_{ij}\bfC_{ii}^0(\tau) + (\bfK_{ij}*\bfC_{jj}^0)(\tau) + (\bfK_{ji}^-*\bfC_{ii}^0)(\tau) + \sum_k (\bfK_{ik}*\bfK_{jk}^-*\bfC_{kk}^0)(\tau),
\end{split}
\end{equation}
where we used $f^-(t) = f(-t)$. \cite{Ostojic:2009} obtained an approximation closely related to Eq.~\eqref{E:first}.
They first obtained the cross-correlation between a pair of neurons which either
receive a common input \emph{or} share a monosynaptic connection. This can be
done using Eq.~\eqref{E:r_approx}, without the need to introduce the mixed process given
in Eq.~\eqref{E:approx}. \cite{Ostojic:2009} then implicitly assumed that the correlations not due to
one of these two submotifs could be disregarded. The correlation between pairs of
cells which were mutually coupled (or were unidirectionally coupled with common input) was approximated by the sum of
correlations introduced by each submotif individually.

 Equation~\eqref{E:first} provides a first approximation to the joint spiking statistics of cells in a recurrent network.  However, it
captures only the effects of direct synaptic connections, represented by the second and third  terms, and 
common input, represented by the last term in Eq.~\eqref{E:first}.  The impact of larger network
structures, such as loops and chains are not captured, although they may significantly impact
cross-correlations~\cite{Roxin:2011,Zhao:2011,Bullmore:2009}. Experimental studies have also shown
that local cortical connectivity may not be fully random~\cite{Song:2005,Oswald:2009,Perin:2011}.  It is 
therefore important to  understand the effects on network architecture on correlations.

To capture the impact of the full network structure,  we propose an iterative approach which accounts for successively larger connectivity patterns in the network~\cite{Rangan:2009-1,Rangan:2009-2}. We again start with $y_i^0(t)$, a realization of a single
spike train in isolation. Successive approximations to the output of cells in a recurrent network are defined by
\begin{equation}\label{E:gen_it_def}
y_i^{n+1}(t) = y_i^0(t) + \sum_j (\bfK_{ij}*[y_j^n-r_j])(t) , \qquad n \geq 0.
\end{equation}

\begin{figure}
\centering
\includegraphics[scale = 1]{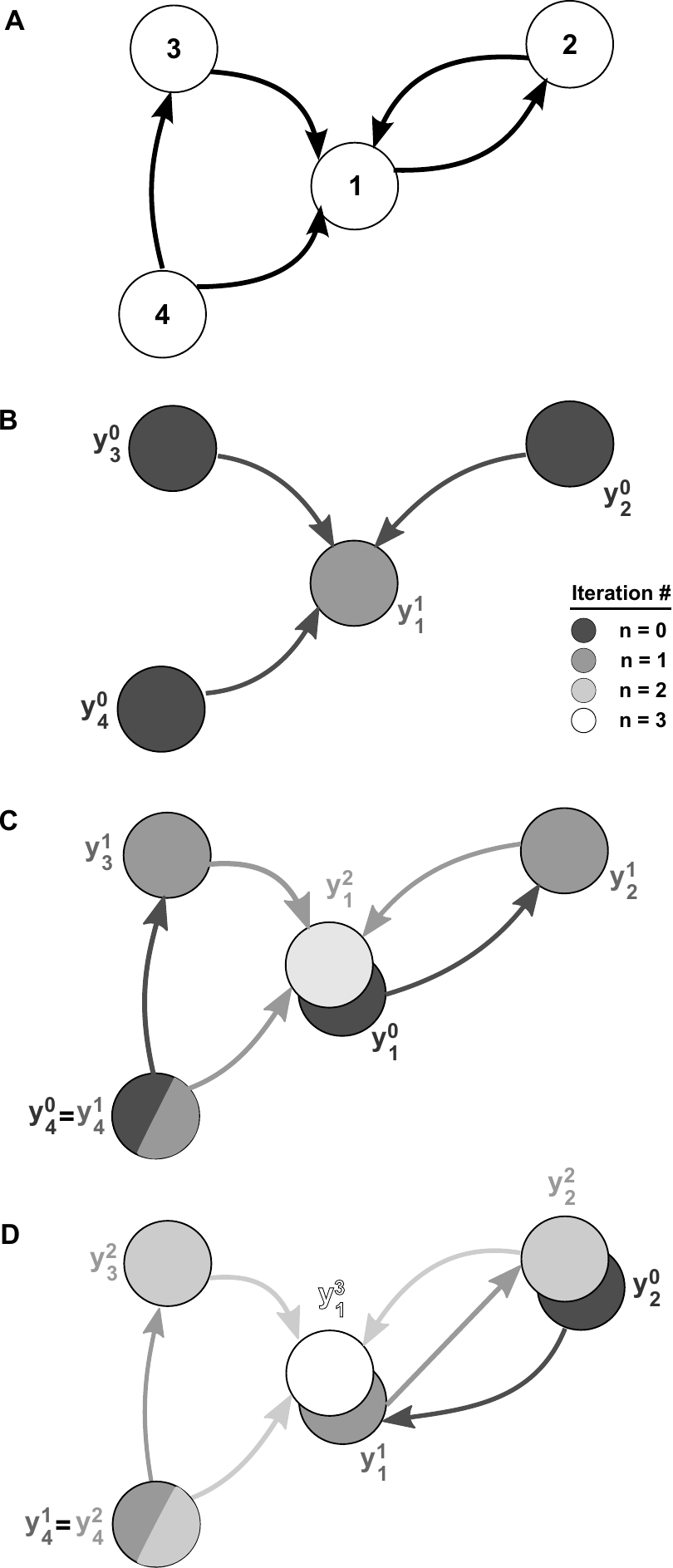}
\caption{ {\bf Iterative construction of the linear approximation to network activity.}
{\bf (A)} An example recurrent network.
{\bf (B)-(D)} A sequence of graphs determines the successive approximations to the output of neuron 1. Processes defined by the same iteration of Eq.~\eqref{E:vector_it} have equal color.
{\bf (B)} In the first iteration of Eq.~\eqref{E:vector_it}, the output of neuron 1 is approximated using
the unperturbed outputs of its neighbors.
{\bf (C)} In the second iteration the results of the first iteration are used to define the inputs to the neuron.  For instance, the process $y_2^1$  depends on the  base process $y_1^0$ which represents the
unperturbed output of neuron 1. Neuron 4 receives no inputs from the rest of the network, and all approximations involve only its unperturbed output, $y_4^0$.
{\bf (D)} Cells 3 and 4 are not part of recurrent paths, and their contributions to the approximation are fixed after the second iteration.  However, the recurrent connection between cells 1 and 2 implies that subsequent  approximations  involve contributions of directed chains of increasing length.
}
\label{F:net_fig}
\end{figure}

To compute the correction to the output of a neuron, in the first iteration we assume that its inputs come from a collection of isolated
cells:  When $n = 1$,  Eq.~\eqref{E:gen_it_def}  takes into account only inputs from
immediate neighbors, treating each as disconnected from the rest of the network. The corrections in
the second iteration are computed using the approximate cell responses obtained from the first iteration.
Thus, with $n = 2$,  Eq.~\eqref{E:gen_it_def}  also accounts for the impact  of next nearest neighbors.  Successive iterations  include the impact of directed chains of increasing length: The isolated output from an independent collection of neurons is filtered
through $n$ stages to produce the corrected response (See Fig.~\ref{F:net_fig}.)

Notation is simplified when this iterative construction is recast in matrix form\footnote{Let $\bfX(t) = [X_{ij}(t)]$ and $\bfY(t) = [Y_{ij}(t)]$ be $n_1 \times n_2$ and $n_2 \times n_3$ matrices of functions, respectively.
We define the convolution of matrices $(\bfX*\bfY)(t)$ to be the $n_1 \times n_3$ matrix of functions with entries defined by
$$
(\bfX*\bfY)_{ij}(t) = \sum_k (X_{ik}*Y_{kj})(t).
$$
Expectations and convolutions commute for matrix convolutions as matrix expectations are taken entry-wise. Each entry of a matrix convolution is a linear combination of scalar convolutions which commute with expectations. Additionally, we adopt the convention that  the zeroth power of the interaction matrix, $\bfK_{ij}^0(t)$, is the diagonal matrix with $\bfK_{ij}^0(t) = \delta(t)$ when $i=j$.  Hence $\bfK_{ij}^0(t)$ acts as the identity matrix under matrix convolution. } to obtain
\begin{equation}\label{E:vector_it}
\begin{split}
\bfy^{n+1}(t) &= \bfy^0(t) + (\bfK*[\bfy^n - \bfr])(t) \\
            &=  \bfy^0(t) + \sum_{k = 1}^{n+1} (\bfK^{(k)}*[\bfy^0 - \bfr])(t),
\end{split}\quad n \geq 0,
\end{equation}
where $\bfy^{n}(t) = [y^n_i(t)]$ and $\bfr = [r_i]$ are length $N$ column vectors, and $\bfK^{(k)}$ represents a $k$-fold matrix convolution of $\bfK$ with itself.

The  $n^{th}$ approximation to the matrix of cross-correlations  can be written in terms of the interaction kernels, $\bfK_{ij},$ and the autocorrelations of the base processes $\bfy^0$ as (See Methods)
\begin{equation}\label{E:matrix_rel_0}
\begin{split}
\bfC_{ij}(\tau)  \approx \bfC^n(\tau) &=  \EVs{(\bfy^n(t+\tau)-\bfr)(\bfy^n(t)-\bfr)^T} \\
&= \sum_{k,l = 0}^{n}( \bfK^{(k)}*\bfC^0*(\bfK^-)^{(lT)})(\tau), \quad n \geq 0,
\end{split}
\end{equation}
where  $\bfK^-(t) = \bfK(-t)$, $\bfX^{(kT)} = (\bfX^{(k)})^T,$ and $\bfX^{(k)}$ is the $k$-fold matrix convolution of $\bfX$ with itself.

If we apply the Fourier transform, 
$
\tilde{f}(\omega) = \mathcal{F}[f(t)](\omega) \equiv \int_{-\infty}^\infty f(t) e^{-2 \pi i \omega t}dt,
$
 to Eq.~\eqref{E:matrix_rel_0}, we find that for each $\omega$,
\begin{equation}\label{E:freq_dom_mat}
\begin{split}
\bfCt^n(\omega) = \EV[\bfyt^{n}(\omega)\bfyt^{n*}(\omega)] &= \sum_{k,l=0}^{n} \bfKt^k(\omega) \EV[ \bfyt^0(\omega)\bfyt^{0*} (\omega)] (\bfKt^{*})^l(\omega) \\
&= \left( \sum_{k=0}^{n}  \bfKt^k(\omega) \right) \EVs{\bfyt^0(\omega)\bfyt^{0*} (\omega)}  \left( \sum_{l=0}^{n} (\bfKt^{*})^l(\omega) \right),
\end{split}
\end{equation}
where $\bfX^*$ denotes the conjugate transpose of $\bfX$.
The zero-mean Fourier transforms $\tilde{y}_i^n$ of the spiking processes $y_i^n$ is defined by $\tilde{y}_i^n = \mathcal{F}[y_i^n-r_i]$, and $\tilde{f} = \mathcal{F}(f)$ for all other quantities.
 
For a suitable matrix norm $||\cdot||$, when $||\bfKt||<1$, we can take the limit $n \rightarrow \infty$ in Eq.~\eqref{E:freq_dom_mat}~\cite{Kato:1995}, to obtain
 an approximation to the full array of cross-spectra
\begin{align}\label{E:freq_dom_lim}
\bfCt(\omega) \approx \bfCt^\infty(\omega) = \lim_{n\rightarrow\infty} \bfCt^n(\omega) = (\bfId-\bfKt(\omega))^{-1}  \bfCt^0 (\omega) (\bfId - \bfKt^*(\omega))^{-1}.
\end{align}
This equation can also be obtained by generalizing the approach of~\cite{Lindner:2005} (also see~\cite{Beck:2011}).
In the limit $n \rightarrow \infty$, directed paths of arbitrary length contribute to the approximation. 
Equation~\eqref{E:freq_dom_lim}   therefore  takes into account the full recurrent structure of the network. We will use the spectral norm $||\cdot||_2$, and assume that in the networks we study $||\bfKt||_2<1$. This condition is confirmed numerically when we use  Eq.~\eqref{E:freq_dom_lim}.

Finally, consider the network response to external signals, $\eta_i(t)$, with zero mean and finite variance. The response of the neurons in the recurrent network can be approximated iteratively by
$$
\bfy^{n+1} = \bfy^0 + \bfK*[\bfy^n-\bfr] + \bfA * \bfeta,
$$
where $\bfA = \text{diag}(A_i)$ and $\bfeta(t) = [\eta_i(t)]$.  External signals and recurrent synaptic inputs are both linearly filtered to approximate a cell's response,  consistent with a generalization of Eq.~\eqref{E:r_approx}. As in Eq.~\eqref{E:matrix_rel_0}, the $n^{th}$  
approximation to the matrix of correlations is
$$
\bfC(\tau) \approx \bfC^n(\tau) = \sum_{k,l=0}^n(\bfK^{(k)} * \bfC^0 * (\bfK^-)^{(lT)})(\tau) + \sum_{k,l=0}^{n-1}(\bfA^{(k)} * \bfC^{\bfeta} * (\bfA^-)^{(lT)})(\tau),
$$
where $\bfC^{\bfeta}(\tau) = \EVs{\bfeta(t+\tau) \bfeta(t)^T}$ is the covariance matrix of the external signals. 
We can again take the Fourier transform and the
limit $n \rightarrow \infty$,
and solve for $\bfCt(\omega)$. If $||\bfKt|| < 1$,
\begin{equation}\label{E:freq_dom_lim_ext}
\bfCt^\infty(\omega) = (\bfI - \bfKt(\omega))^{-1}(\bfCt^0(\omega) + \bfAt(\omega) \bfCt^{\bfeta}(\omega) \bfAt^*(\omega))(\bfI - \bfKt^*(\omega))^{-1}.
\end{equation}
When the signals comprising $\bfeta$ are white (and possibly correlated) corrections must be made to account for the change in spectrum and response properties of the isolated cells~\cite{Lindner:2005,delaRocha:2007, Vilela:2009} (See Methods). 
 
We note that Eq.~\eqref{E:vector_it}, which is the basis of our iterative approach, provides an approximation to the network's output which is of higher than first order in connection strength.  This may seem at odds with a theory that
provides a linear correction to a cell's response, \emph{cf.} Eq.~\eqref{E:r_approx}.
However, Eq.~\eqref{E:vector_it} does not  capture nonlinear corrections to the response
of individual cells, as the output of each cell is determined linearly  from its
input.  It is the input that can contain terms of any order in connection strength stemming
from  directed paths of different lengths through the network.


\section*{Results}

We  use the theoretical  framework developed above to analyze the statistical structure of  the spiking activity in a network of IF neurons described by Eq.~\eqref{E:main}.  We first show that the cross-correlation functions between
cells in two small networks can be studied in terms of contributions from directed paths through the network.   We
use a similar approach to understand the structure of correlations in larger all--to--all and random networks.  
We show that in networks where inhibition and excitation are tuned for exact balance, only local interactions contribute to correlations.  When such balance is broken by a relative elevation of inhibition, the result may be increased synchrony in the network.  The theory also allows us to obtain averages of cross-correlation functions conditioned on connectivity between pairs of cells in random networks.  Such averages
can provide a tractable yet accurate description of the joint statistics of spiking in these networks.  

  The correlation structure is determined by the response properties of cells together with  synaptic dynamics and network architecture.     Network interactions are described by
the matrix of synaptic filters, $\bfJ$, given in Eq.~\eqref{E:synaptic}, while 
the response of cell $i$ to an input is approximated using its linear response kernel $A_i$.  
Synaptic dynamics, architecture, and cell responses are all combined in the matrix
$\bfK$, where $\bfK_{ij}$ describes the response of
cell $i$ to an input from cell $j$ (See Eq.~\eqref{E:main}).   The correlation structure of network activity is approximated in 
Eq.~\eqref{E:freq_dom_lim} using the Fourier transforms of the interaction matrix, $\bfK$, and the matrix of unperturbed autocorrelations $\bfC^0$.

\subsection*{Statistics of the response of microcircuits}

\begin{figure}[htb!]
\centering
\includegraphics[scale = 0.9]{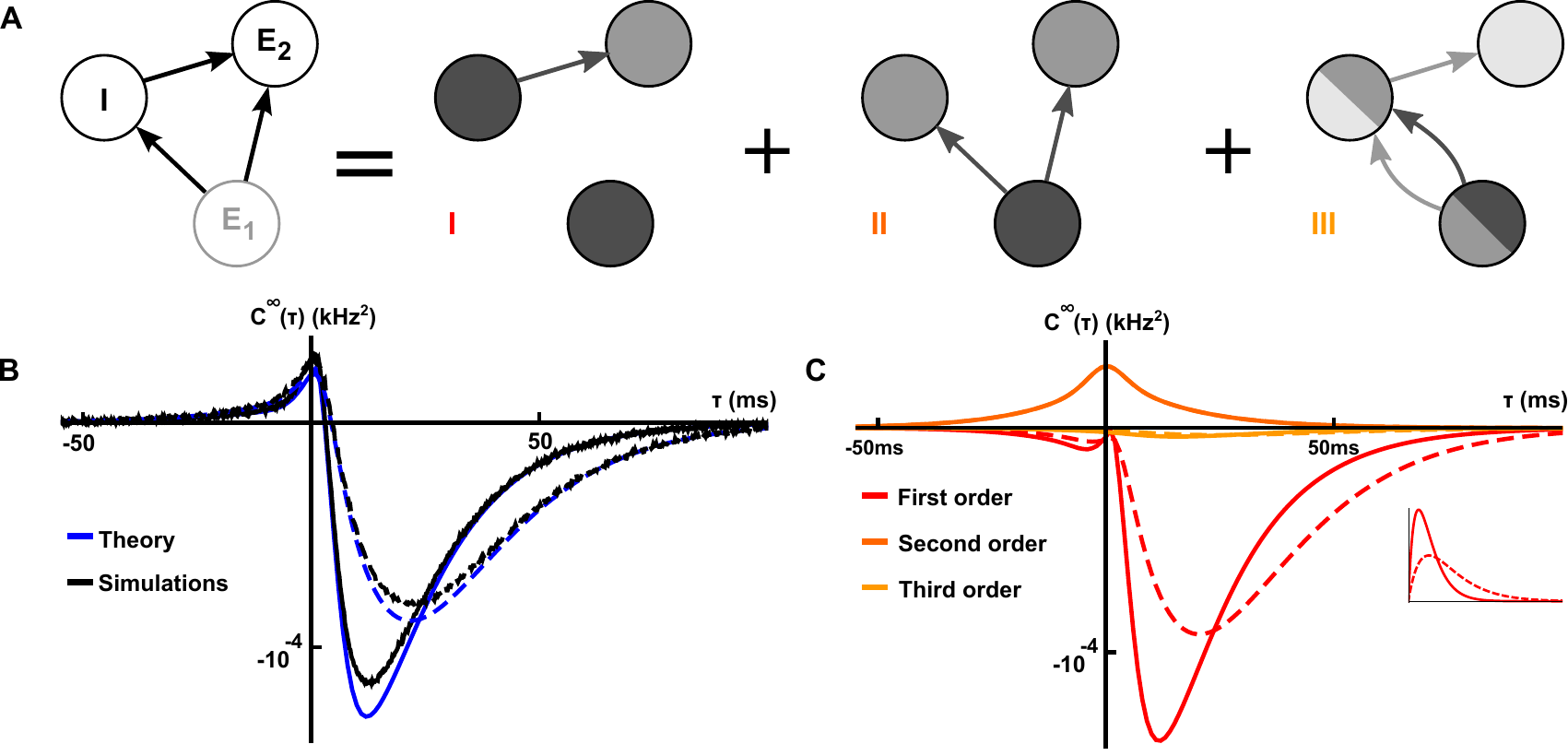}
\caption{{\bf The relation between correlation structure and response statistics in a feed-forward inhibitory microcircuit.} 
{\bf (A)}  The FFI circuit (left) can be decomposed into three
submotifs.   Equation~\eqref{E:ffi_ei_cs} shows that each submotif provides a specific contribution to the cross-correlation between 
cells $E_2$ and $I$.
{\bf (B)}  Comparison of the theoretical prediction with the numerically computed cross-correlation between 
cells $E_2$ and $I$.  Results are shown for two different values of the inhibitory time constant,$\tau_I$
($\tau_I = 5$ms, solid line, $\tau_I = 10$ms, dashed line).
{\bf (C)} The contributions of the different submotifs  in panel  A are shown  for both $\tau_I = 5$ms (solid) and $\tau_I = 10$ms (dashed). Inset shows the corresponding change in the inhibitory synaptic filter.
The present color scheme is used  in subsequent figures. Connection strengths were $\pm 40 \ \text{mV} \cdot \text{ms}$ for excitatory and inhibitory connections. In each case, the long window correlation coefficient $\bfrho_{E_2I}$ between the two cells was $\approx -0.18$.
}
\label{F:ffi_fig}
\end{figure}

 We first consider a pair of simple  microcircuits to highlight some of the  features of the theory. We start with
the three cell model of feed-forward inhibition shown in Fig.~\ref{F:ffi_fig}A~\cite{Kremkow:2010}. The  interaction matrix, $\bfKt(\omega)$, has  the  form 
$$
\bfKt(\omega) = \left(\begin{matrix} 0 & 0 & 0\\ \Kt_{E_2E_1}(\omega) & 0 & \Kt_{E_2I}(\omega)\\ \Kt_{IE_1}(\omega) & 0 & 0 \end{matrix}\right),
$$
where cells are indexed in the order $E_1,E_2,I$.  To simplify notation,  we 
omit the dependence of $\bfKt(\omega)$ and other spectral quantities on $\omega$. 

Note that $\bfKt$ is nilpotent, and the inverse of $(\bfI - \bfKt)$ may be expressed as
\begin{equation}\label{E:ffi_inv}
(\bfI-\bfKt)^{-1} = (\bfI + \bfKt + \bfKt^2) = \left(\begin{matrix} 1 & 0 & 0 \\ \Kt_{E_2E_1} + \Kt_{E_2I}\Kt_{IE_1} & 1 & \Kt_{E_2I} \\ \Kt_{IE_1} & 0 & 1 \end{matrix}\right).
\end{equation}
Substituting Eq.~\eqref{E:ffi_inv} into Eq.~\eqref{E:freq_dom_lim}  yields an approximation to the matrix of cross-spectra.  For instance,
\begin{align}
\Ct_{E_2I}^\infty &= \Kt_{IE_2}\Ct_I^0 + \Kt_{E_2E_1}\Kt_{IE_1}^*\Ct_{E_1}^0  + \Kt_{E_2I}|\Kt_{IE_1}|^2 \Ct_{E_1}^0 \notag\\
&= \underbrace{(\At_{E_2}\Jt_{E_2I} )\Ct_I^0}_{\rm I} + \underbrace{(\At_{E_2}\Jt_{E_2E_1})(\At_I\Jt_{IE_1})^*\Ct_{E_1}^0}_{\rm II} \label{E:ffi_ei_cs}\\
&\qquad\qquad +  \underbrace{(\At_{E_2}\Jt_{E_2E_1})|\At_I\Jt_{IE_1}|^2\Ct_{E_1}^0}_{\rm III}. \notag
\end{align}
Figure~\ref{F:ffi_fig}B shows that these approximations closely match numerically obtained cross-correlations. $\Ct_X^0$ is the uncoupled power spectrum
for cell $X$.

Equation~\eqref{E:ffi_ei_cs} gives insight into how the joint response of cells in this circuit is shaped by the features of the network.  The three terms in Eq.~\eqref{E:ffi_ei_cs}  are directly related to the architecture of the microcircuit:
Term I represents the correlating effect of the direct input to cell $E_2$ from cell $I$.  Term II
captures the  effect of the common input from cell $E_1$. Finally, term III represents the interaction of
the indirect input from $E_1$ to $E_2$ through $I$ with the input from $E_1$ to $I$ (See Fig.~\ref{F:ffi_fig}C).
A change in any single parameter may affect multiple terms.  However, the individual contributions of all three terms
are apparent. 
    
To illustrate the impact of synaptic properties on the cross-correlation between cells $E_2$ and $I$ we varied the inhibitory time constant, $\tau_I$  (See Fig.~\ref{F:ffi_fig}B and C).  Such a change is primarily reflected in the shape of the first order term, I:  Multiplication by  $\Jt_{E_2I}$ 
is equivalent to convolution with the  inhibitory synaptic filter, $J_{E_2I}$.  The shape of this filter is determined by
$\tau_I$ (See Eq.~\eqref{E:synaptic}), and a shorter time constant leads to a tighter timing dependency between
the spikes of the two cells~\cite{Veredas:2005,Kirkwood:1979,Ostojic:2009,Fetz:1983,Herrmann:2001}.  In particular, Ostojic \emph{et al.} made similar observations using a related  approximation.
In the FFI circuit, the first and second order terms, I and II, are dominant (red and dark orange, Fig.~\ref{F:ffi_fig}B).  The relative magnitude of the third order term, III (light orange, Fig.~\ref{F:ffi_fig}B), is small.  The next example shows that even in a simple recurrent circuit, terms of order higher than two may be significant.

\begin{figure}[p]
\centering
\includegraphics[scale = 0.9]{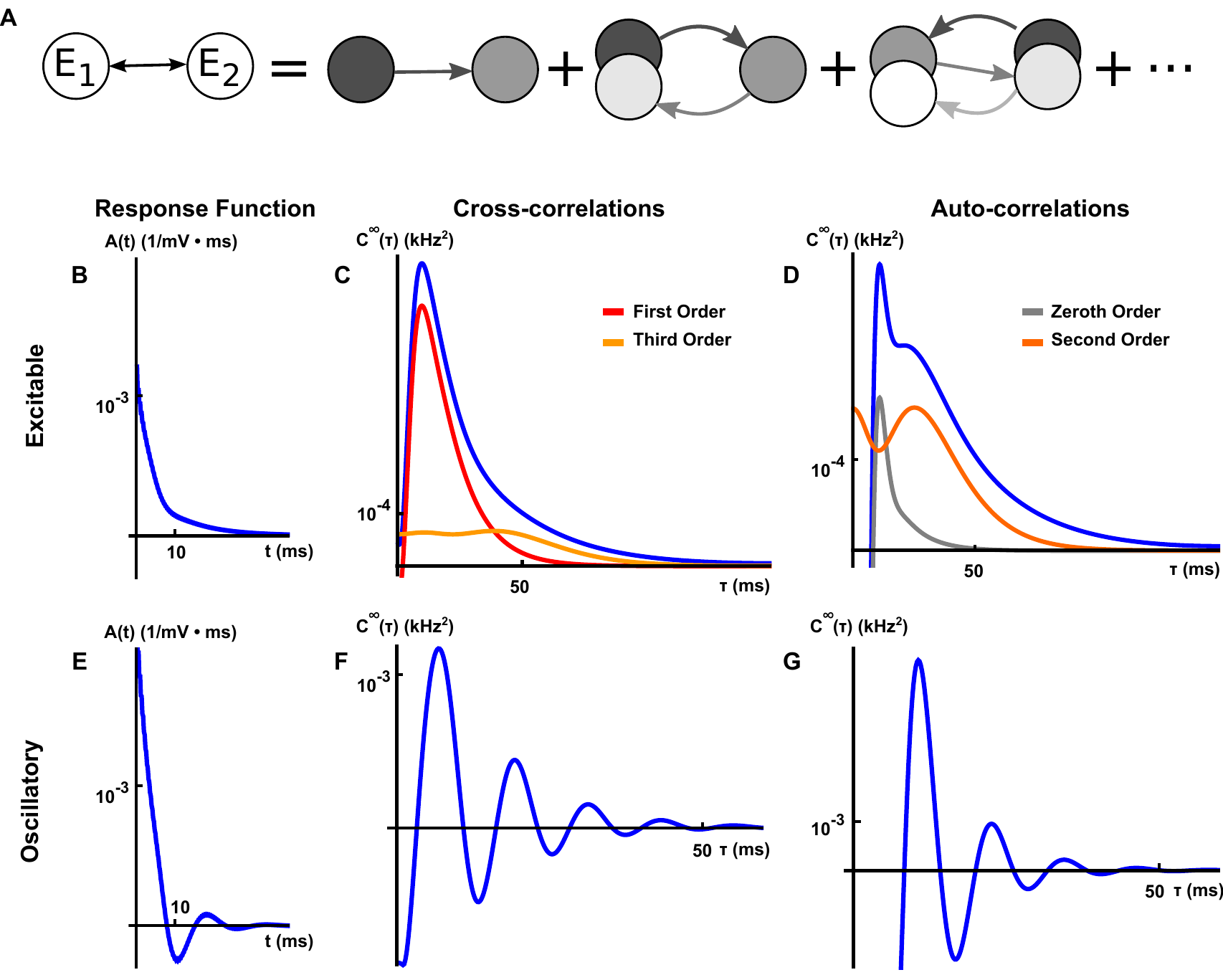}
\caption{
{\bf The relation between correlation structure and response statistics for two bidirectionally coupled, excitatory cells.} 
{\bf (A)} The cross-correlation between the two  cells
 can be represented in terms of contributions from an infinite sequence
of submotifs (See Eq.~\eqref{E:bidir_cross}). Though we show only a few ``chain" motifs in one direction, one should note that
there will also be contributions to the cross-correlation from chain motifs in the reverse direction in addition to indirect common input
motifs (See the discussion of Fig.~\ref{F:motif_fig}).
{\bf (B), (E)} Linear response kernels  in the excitable (B) and oscillatory (E) regimes.
{\bf (C), (F)} The cross-correlation function with  first and third order contributions computed using Eq.~\eqref{E:bidir_sol} in the excitable (C) and oscillatory (F) regimes.
{\bf (D), (G)} The auto-correlation function with zeroth and second order contributions computed using Eq.~\eqref{E:bidir_sol} in the excitable (D) and oscillatory (G) regimes.
In the oscillatory regime, higher order contributions were small relative to first order contributions and are therefore not shown.
The network's symmetry implies that cross-correlations are symmetric, and we only show them 
for positive times.  All cross-correlations are nearly indistinguishable from those obtained 
from simulations (See SI Fig. 3 for comparisons with simulations).  Connection strengths were $40 \ \text{mV}\cdot\text{ms}$. The long window correlation coefficient $\bfrho_{E_1E_2}(\infty)$ between the two cells was $\approx 0.8$ in the excitable regime and $\approx 0.5$ in the oscillatory regime. The ISI CV was approximately 0.98 for neurons in the excitable regime and 0.31 for neurons in the oscillatory regime.}
\label{F:ee_fig}
\end{figure}

More generally, the interaction matrices, $\bfKt$, of recurrent networks  are not nilpotent.
Consider two reciprocally coupled excitatory cells, $E_1$ and $E_2$ (See  Fig.~\ref{F:ee_fig}A, left). In this case, $$
\bfKt = \left(\begin{matrix} 0 & \Kt_{E_1E_2} \\ \Kt_{E_2E_1} & 0 \end{matrix} \right)
$$
so that
$$
(\bfI - \bfKt)^{-1} = \frac{1}{1 - \Kt_{E_1E_2} \Kt_{E_2E_1}}(\bfI + \bfKt).
$$
Equation~\eqref{E:freq_dom_lim} gives the following approximation to the matrix of cross-spectra
\begin{equation}\label{E:bidir_sol}
\begin{split}
\bfCt^\infty &= \frac{1}{|1 - \Kt_{E_1E_2} \Kt_{E_2E_1}|^2}(\bfI + \bfKt)\left(\begin{matrix} \Ct_{E_1}^0 & 0 \\ 0 & \Ct_{E_2}^0 \end{matrix}\right)(\bfI + \bfKt^*) \\
&=  \frac{1}{|1 - \Kt_{E_1E_2} \Kt_{E_2E_1}|^2} \left(\begin{matrix}\Ct_{E_1}^0 + |\Kt_{E_1E_2}|^2\Ct_{E_2}^0 & \Kt_{E_2E_1}^*\Ct_{E_1}^0 + \Kt_{E_1E_2}\Ct_{E_2}^0 \\ K_{E_2E_1}\Ct_{E_1}^0 + K_{E_1E_2}^*\Ct_{E_2}^0 & \Ct_{E_2}^0 + |K_{E_2E_1}|^2\Ct_{E_1}^0 \end{matrix}\right).
\end{split}
\end{equation}
In contrast to the previous example, this approximation does not terminate at finite order in interaction strength. After expanding, the cross-spectrum between cells $E_1$ and $E_2$ is approximated by
\begin{equation}\label{E:bidir_cross}
\Ct_{E_1E_2}^\infty = \sum_{k,l=0}^\infty (\Kt_{E_1E_2}\Kt_{E_2E_1})^k (\Kt_{E_1E_2}^*\Kt_{E_2E_1}^*)^l ( \Kt_{E_2E_1}^*\Ct_{E_1}^0 + \Kt_{E_1E_2}\Ct_{E_2}^0).
\end{equation}
Directed paths beginning at $E_1$ and ending at $E_2$ (or vice-versa) are of odd length. Hence, this approximation  contains only odd powers of the kernels $\Kt_{E_iE_j}$, each corresponding  to a directed path from one cell to the other. Likewise, the approximate power spectra  contain only even powers of the kernels corresponding to directed paths that connect a cell to itself (See Fig.~\ref{F:ee_fig}A).

The contributions of different sub-motifs to the cross- and auto-correlations are shown in Figs.~\ref{F:ee_fig}C, D when the isolated cells are in a near-threshold excitable state ($\text{CV} \approx 0.98$).  The auto-correlations are significantly affected by network interactions.  We also note that chains of length two and three (the second and third submotifs in Fig.~\ref{F:ee_fig}A) provide significant contributions.     
Earlier approximations
do not capture such corrections~\cite{Ostojic:2009}.

 The operating point of a cell is set by its parameters ($\tau_i, E_{L,i}$, etc.) and the statistics of its input ($E_i, \sigma_i$). 
A change in operating point can significantly change a cell's  response to an input.
Using linear response theory, these changes  are reflected in the response functions $A_i$, and the power spectra of the
 isolated cells, $\bfCt^0$. To highlight the role that the operating point plays in the approximation of the correlation structure given by Eq.~\eqref{E:freq_dom_lim}, we elevated the 
mean and decreased the variance of background noise by increasing $E_i$ and decreasing $\sigma_i$ in  Eq.~\eqref{E:main}.   With the chosen parameters the isolated cells are in a super-threshold, low noise regime and fire nearly periodically ($\text{CV} \approx 0.31$).
After the cells are coupled,  this oscillatory behavior is reflected in the cross- and auto-correlations where the dominant contributions are due to first and zeroth order terms, respectively (See Figs.~\ref{F:ee_fig}F,G).

\begin{figure}[htb!]
\centering
\includegraphics[scale = 0.9]{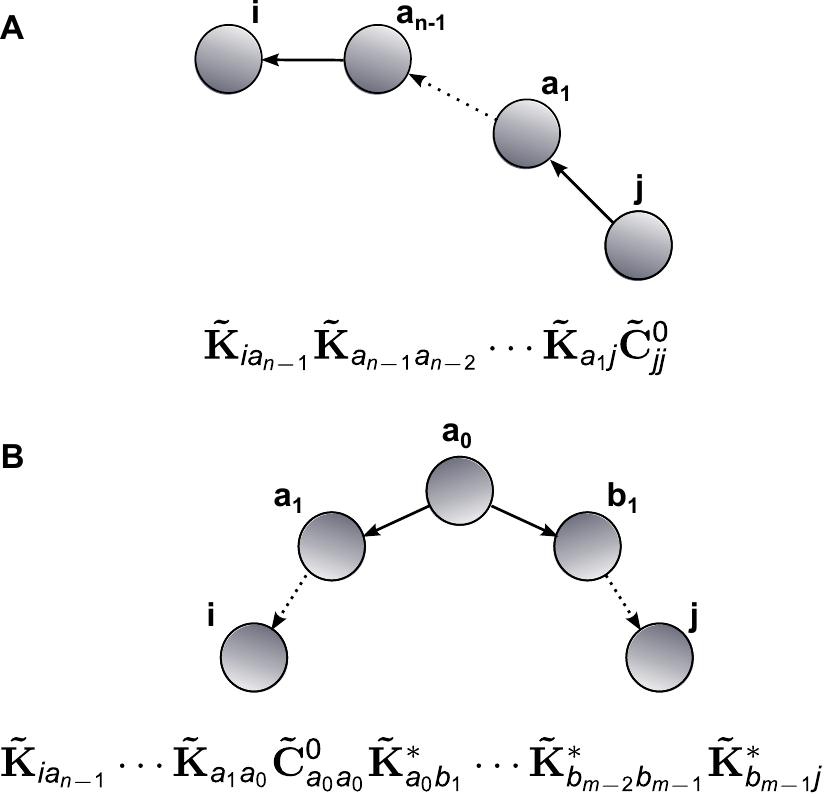}
\caption{
{\bf The motifs giving rise to terms in the expansion of Eq.~\eqref{E:freq_dom_lim}.} 
{\bf (A)} Terms containing only unconjugated (or only conjugated) interaction kernels $\bfKt_{ab}$  correspond to directed chains.
{\bf (B)} Terms containing both unconjugated and conjugated interaction kernels $\bfKt_{ab}$ correspond to direct or indirect common input motifs.
}
\label{F:motif_fig}
\end{figure}

  \paragraph{Orders of coupling interactions:} It is often useful to expand Eq.~\eqref{E:freq_dom_lim} in terms of powers of $\bfKt$~\cite{Pernice:2011}.  The term $\bfKt^n \bfCt^0 (\bfKt^*)^m$ in the expansion is said to be of \emph{order} $n+m$. Equivalently, in the expansion of $\bfCt^\infty_{ij}$,  the order of a term refers to the sum of the powers of all constituent interaction kernels $\bfKt_{ab}$. We can also associate a particular connectivity submotif with each term. In particular, $n^{\text{th}}$ order terms of the form
 $$
 \bfKt_{ia_{n-1}}\bfKt_{a_{n-1}a_{n-2}}\cdots\bfKt_{a_1j}\bfCt^0_{jj} 
 $$
are associated with a directed path $j \rightarrow a_1 \rightarrow \cdots \rightarrow a_{n-2} \rightarrow a_{n-1} \rightarrow i$ from cell $j$ to cell $i$. Similarly, the term $\bfCt^0_{ii}  \bfKt^*_{ia_1}\cdots\bfKt^*_{a_{n-2}a_{n-1}}\bfKt^*_{a_{n-1}j}$ corresponds to a $n$-step path from cell $i$ to cell $j$. An $(n+m)^{\text{th}}$ order term of the form
$$
 \bfKt_{ia_{n-1}}\bfKt_{a_{n-1}a_{n-2}}\cdots\bfKt_{a_1a_0}\bfCt^0_{a_0a_0} \bfKt_{a_0b_1}^*\cdots\bfKt_{b_{m-2}b_{m-1}}^*\bfKt_{b_{m-1}j}^*
 $$
represents the effects of an indirect common input $n$ steps removed from cell $i$ and $m$ steps removed from cell $j$.  This corresponds to a submotif of the form  $i \leftarrow a_{n-1} \leftarrow \cdots \leftarrow a_0 \rightarrow b_1 \rightarrow \cdots \rightarrow b_{n-1}\rightarrow j$ consisting of two branches originating at cell $a_0$.  (See Fig.~\ref{F:motif_fig}, and also Fig.~\ref{F:all_fig}A and the discussion around Eqs.~(\ref{E:ffi_ei_cs},\ref{E:bidir_cross}).)


\subsection*{Statistics of the response of large networks}

The full power of the present approach  becomes evident when analyzing the activity of larger
networks. 
We again illustrate the theory using several examples. In networks where inhibition and excitation are tuned to be precisely balanced, the theory shows that only local interactions contribute to correlations.  
When this balance is broken, terms corresponding to longer paths through the network 
shape the cross-correlation functions.   One consequence is that a relative increase in inhibition can
lead to elevated network synchrony.  We also show how to obtain tractable and
accurate approximation of the average correlation structure in random networks.

\paragraph{A symmetric, all--to--all network of excitatory and inhibitory neurons}

We begin with an all--to--all coupled network of $N$ identical cells. Of these cells, $N_E$ make excitatory, and $N_I$ make inhibitory synaptic connections.    The excitatory cells are assigned indices $1,\ldots,N_E,$ and the  inhibitory cells indices $N_E+1,\ldots,N$.  All excitatory (inhibitory) synapses have weight $W_E = \frac{G_E}{N_E}$ ($W_I = \frac{G_I}{N_I}$), and timescale $\tau_E$ ($\tau_I$). The interaction matrix $\bfKt$  may then be written in block form,
$$
\bfKt = \At  \bfJt, \qquad \text{where} \qquad \bfJt = 
\left(\begin{matrix} \Jt_{E} \bfone_{N_EN_E} & 
\Jt_I \bfone_{N_EN_I} \\ 
\Jt_E \bfone_{N_IN_E} &
\Jt_I\bfone_{N_IN_I} \end{matrix}\right).
$$
Here ${\bf 1}_{N_1N_2}$ is the $N_1\times N_2$ matrix of ones, $\Jt_X$ is the weighted synaptic kernel for cells of class $X$ (assumed  identical within each class), and $\At$ is the susceptibility function for each cell in the network. Although the effect of autaptic connections (those from a cell to itself)  is negligible (See SI Fig. 2, their inclusion significantly simplifies the resulting expressions.

\begin{figure}
\centering
\includegraphics[scale=0.9]{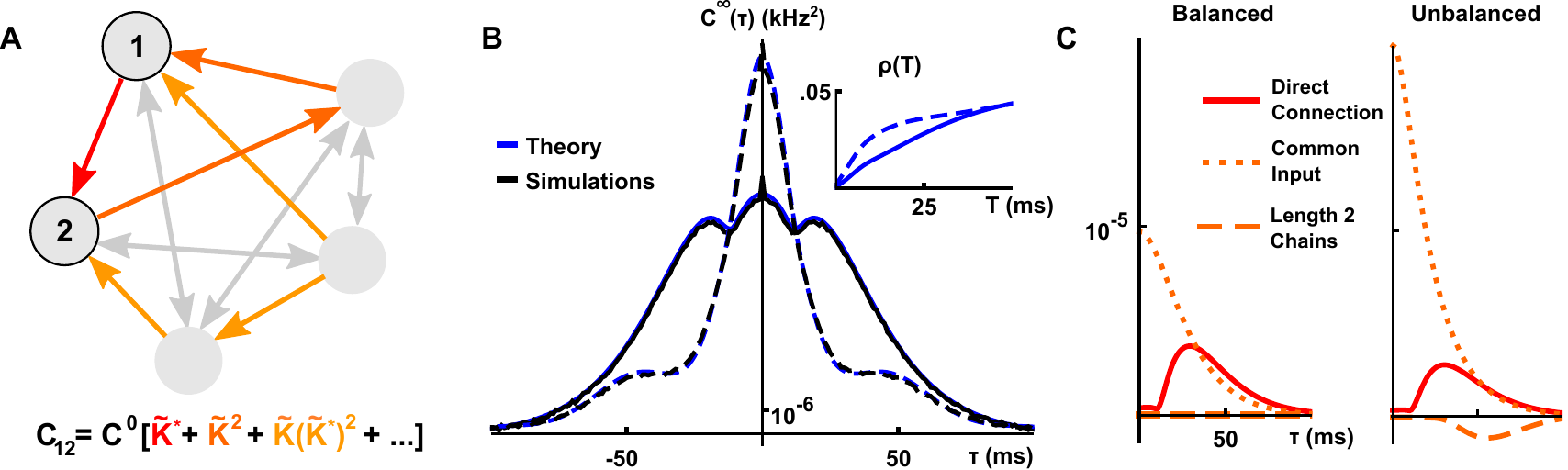}
\caption{ {\bf All--to--all networks and the importance of higher order motifs.}
{\bf (A)} Some of the submotifs contributing to correlations in the all--to--all network.
{\bf (B)} Cross-correlations between two excitatory cells in an all--to-all network ($N_E = 80, N_I = 20$) obtained 
 using Eq.~\eqref{E:alltoall_c} 
 (Solid -- precisely tuned network with $\mut \equiv 0$ [$G_E = 175 \text{ mV$\cdot$ms}, G_I = -(N_E/N_I)G_E = -700  \text{ mV$\cdot$ms}, \tau_E = \tau_I = 10\text{ ms}$], 
 dashed -- non-precisely tuned network with $\mut \neq 0$  [$G_E = 210  \text{ mV$\cdot$ms}, G_I = -1050  \text{ mV$\cdot$ms}, \tau_E = 10 \text{ ms}, \tau_I = 5 \text{ ms}$]).
{\bf (C)} Comparison of first and second order contributions to the cross-correlation function in panel A in the precisely tuned (left) and non-precisely tuned (right) network. In both cases, the long window correlation coefficient $\rho(\infty)$ was fixed at 0.05.
}
\label{F:all_fig}
\end{figure}

We define
$\mut_E = N_E\Jt_E, \mut_I =N_I\Jt_I$, and $\mut = \mut_E + \mut_I$. Using induction, we can show that 
$$
\bfKt^k = \At^k \mut^{k-1} \bfJt.
$$
Direct matrix multiplication yields 
$$
\bfJt\bfJt^* = \mut_c \bfone_{NN}\quad\text{where}\quad \mut_c = N_E|\Jt_E|^2 + N_I |\Jt_I|^2,
$$
which allows us to calculate the powers $\bfKt^k\bfKt^{l*}$ when $k,l \neq 0$,
$$
\bfKt^k\bfKt^{l*} = \At^k (\At^{*})^l \mut^{k-1}(\mut^*)^{l-1}\mut_c \bfone_{NN}.
$$
An application of Eq.~\eqref{E:freq_dom_lim} then gives an approximation to the matrix of cross-spectra:
\begin{equation}\label{E:alltoall_c}
\bfCt^\infty = \Ct^0 \sum_{k,l = 0}^\infty \bfKt^k\bfKt^{l*} = \Ct^0 \left[  \left(\frac{\At}{1 - \At\mut}\right) \bfJt + \left(\frac{\At}{1 - \At\mut}\right)^*\bfJt^* + \left|\frac{\At}{1-\At\mut} \right|^2 \mut_c \bfone_{NN}+ \bfI_N  \right]
\end{equation}
The cross-spectrum between two cells in the network is therefore given by
\begin{equation}\label{E:alltoall_pair}
[\bfCt_{ij}^\infty]_{i \in X, j \in Y} =   \Ct^0\left[ \ \left(\frac{\At}{1 - \At\mut}\right)\frac{\mut_Y}{N_Y} + \left(\frac{\At}{1 - \At\mut}\right)^*\frac{\mut_X^*}{N_X} + \left|\frac{\At}{1-\At\mut} \right|^2 \mut_c + \delta_{ij} \right], 
\end{equation}
where  $X \in \{E,I\}$. In Eq.~\eqref{E:alltoall_pair} the first two terms represent the effects of all unidirectional chains  originating at cell $j$ and terminating at cell $i$, and vice versa.   The third term represents the effects of direct and indirect common inputs to the two neurons. In Fig.~\ref{F:all_fig}A, we highlight a few of these contributing motifs.

Interestingly, when excitation and inhibition are tuned for precise balance (so that $\mut = \mut_E + \mut_I =  0$), Eq.~\eqref{E:alltoall_pair} reduces to
\begin{equation}\label{E:alltoall_pair_bal}
[\bfCt^\infty]_{i \in X, j \in Y} =   \Ct^0\left[  \At\frac{\mut_Y}{N_Y} + \At^*\frac{\mut_X^*}{N_X} + |\At|^2 \mut_c + \delta_{ij} \right].
\end{equation}
Effects of direct connections between the cells are captured by the first two terms, while those of direct common inputs to the pair are captured by the third term. Contributions from other paths do not appear. In other words, \emph{in the precisely balanced case only local interactions contribute to correlations.}

To understand this cancelation intuitively, consider the contribution of directed chains originating at a given excitatory neuron, $j$.   For $\tau>0$, the cross-correlation function, $\bfC_{ij}(\tau)$,
is determined by the change in firing rate of cell $i$ at time $\tau$ given a spike in cell $j$ at time 0. 
By the symmetry of the all--to--all connectivity and stationarity, the firing of cell $j$ has an equal probability of eliciting a spike in any excitatory or inhibitory cell in the network.  Due to the precise synaptic balance, the postsynaptic current generated by the elicited spikes in the excitatory population will cancel the postsynaptic current due to elicited spikes  in the inhibitory population on average. The contribution of other motifs cancel in a similar way.

In Fig.~\ref{F:all_fig}B, we show the impact of breaking this excitatory-inhibitory balance  on cross-correlation functions. We increased the strength and speed of the inhibitory synapses relative to excitatory synapses, while
holding constant, for sake of comparison, the long window correlation coefficients, $\rho(\infty)$. at $\approx 0.05$  Moreover, the degree of network synchrony, characterized by the short window correlation coefficients,  is increased  (See Fig.~\ref{F:all_fig}B inset).  
Intuitively,  a spike in one of the excitatory cells  transiently increases the likelihood of spiking in all other cells in the network. Since inhibition in the network is stronger and faster than excitation, these additional spikes will transiently decrease the likelihood of spiking in twice removed cells.

Linear response theory allows us to confirm this heuristic observation, and quantify the impact of the imbalance on second order statistics. Expanding Eq.~\eqref{E:alltoall_pair} for two excitatory cells to second order in coupling strength, we find
\begin{equation}\label{E:alltoall_eepair_exp}
\Ct_{E_iE_j}^\infty = \Ct^0\left[\At \frac{\mut_E}{N_E} +\At^*\frac{\mut_E^*}{N_E} + |\At|^2\mut_c  +
\underline{\At^2\mut \frac{\mut_E}{N_E}   +(\At^*)^2\mut^*\frac{\mut_E^*}{N_E}} + \delta_{ij} \right] + \mathcal{O}(||\bfKt||^3).
\end{equation}
The complete cancellation between contributions of chains involving excitatory and inhibitory cells no longer takes place, and the two underlined terms appear as a consequence (compare with Eq.~\eqref{E:alltoall_pair_bal}).  These underlined terms capture the effects of all length two chains between cells $E_i$ or $E_j$, starting at one  and terminating at the other.  The relative strengthening of inhibition implies that  chains of length two provide a negative contribution to the cross-correlation function at short times (\emph{cf.}~\cite{Vreeswijk:1994}). Additionally, the impact of direct common input to cells $E_i$ and $E_j$ on correlations is both larger in magnitude (because we increased the strength of both connection types) and sharper (the faster inhibitory time constant means common inhibitory inputs induces sharper correlations). 
These changes are reflected in the second order term $|\At|^2\mut_c$ in Eq.~\eqref{E:alltoall_eepair_exp}. 

In sum, unbalancing excitatory and inhibitory connections via stronger, faster inhibitory synapses enhances synchrony, moving a greater proportion of the covariance mass closer to $\tau = 0$ (See Fig.~\ref{F:all_fig}B).  To illustrate this effect in terms of underlying connectivity motifs, we show the contributions of length two chains and common input in both the precisely tuned and non-precisely tuned cases  in Fig.~\ref{F:all_fig}C.   A similar  approach would allow us to understand the impact of a wide range of changes in cellular or synaptic dynamics on the structure of correlations across networks.

%
%

\paragraph{Random, fixed in-degree networks of homogeneous excitatory and inhibitory neurons}

Connectivity in cortical neuronal networks is typically sparse, and connection probabilities can follow distinct rules depending on area and layer~\cite{Shepherd:1991}.  The present theory allows us to
consider arbitrary architectures, as we now illustrate.

We consider a randomly connected network of  $N_E$ excitatory and $N_I$ inhibitory cells coupled with probability $p$. To simplify the analysis, every cell receives exactly $pN_E$ excitatory and $pN_I$ inhibitory inputs. Thus, having fixed  in-degree,  each cell receives an identical level of mean synaptic input. In addition, we continue to assume that cells are identical.  Therefore, the response of each cell in the network is described by the same  linear response kernel.     The excitatory and inhibitory connection strengths  are $G_E/(pN_E)$ and $G_I/(pN_I)$, respectively.  The timescales of excitation and inhibition may differ, but are again identical for cells within each class.

\begin{figure}
\centering
\includegraphics[scale=0.9]{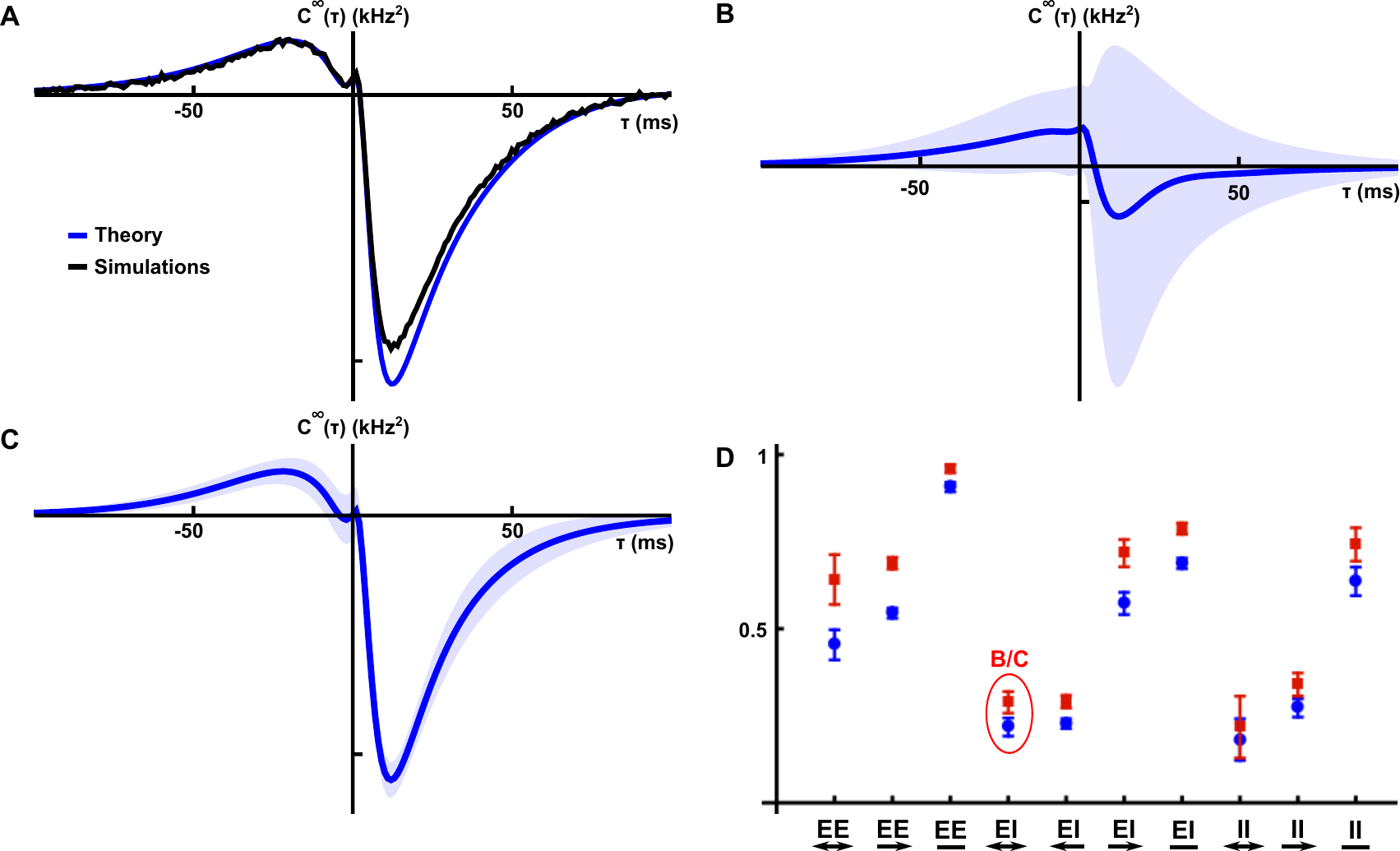}
\caption{
{\bf Correlations in random,  fixed in-degree networks.}
{\bf (A)} A comparison of numerically obtained excitatory-inhibitory cross-correlations to the approximation given by Eq.~\eqref{E:random_foc_av}.
{\bf (B)} Mean and standard deviation for the distribution of correlation functions  for  excitatory-inhibitory pairs of cells. (Solid line -- mean cross-correlation, shaded area -- one standard deviation from the mean, calculated using bootstrapping in a single network realization). 
{\bf (C)} Mean and standard deviation for the distribution of cross-correlation functions conditioned on cell type \emph{and} first order connectivity for a reciprocally coupled excitatory-inhibitory pair of cells. (Solid line -- mean  cross-correlation function, shaded area -- one standard deviation from the mean found by bootstrapping). 
{\bf (D)} Average reduction in $L^2$ error between cross-correlation functions and their respective first-order conditioned averages, relative to the error between the cross-correlations and their cell-type averages. Blue circles give results for a precisely tuned network, and red squares for a network with stronger, faster inhibition. Error bars indicate two standard errors above and below the mean.
$G_E,G_I,\tau_E,\tau_I$ for panels A-C are as in the precisely tuned network of Fig.~\ref{F:all_fig}, and the two networks of panel D are as in the networks of the same figure.
}
\label{F:rand_fig}
\end{figure}

The approximation of network correlations (Eq.~\eqref{E:freq_dom_lim}) depends on the realization of the connectivity matrix.   For a fixed realization, the underlying equations can be solved numerically to approximate the correlation structure (See Fig.~\ref{F:rand_fig}A). However,  the cross-correlation between a pair of cells of given types has a form which is easy to analyze when only leading order terms in $1/N$ are retained.


Specifically, the average cross-spectrum for two cells of given types is (See SI Section 1)
\begin{equation}\label{E:random_ct_av}
\EVb{ \bfCt_{ij}^\infty}_{i\in X, j \in Y} = \Ct^0 \left[\left(\frac{\At}{1 - \At\mut}\right) \frac{\mut_Y}{N_Y}+ \left(\frac{\At}{1 - \At\mut}\right)^*\frac{\mut_X^*}{N_X} + \left|\frac{\At}{1-\At\mut} \right|^2 \mut_c \right] + \mathcal{O}(1/N^2),
\end{equation}
when $i \neq j$. This shows that, to leading order in $1/N$, the mean cross-spectrum between two cells in given classes equals that in the all--to--all network (see Eq.~\eqref{E:alltoall_pair}).  
Therefore our previous discussion relating network architecture
to the shape of cross-correlations in the all--to--all network extends to the 
average correlation structure in the random network for large $N$. 


\cite{Pernice:2011} derived similar expressions for the correlation functions in networks of interacting \emph{Hawkes processes}~\cite{Hawkes:1971-1,Hawkes:1971-2} by assuming either the network is regular (i.e., both in- and out-degrees are fixed) or has a sufficiently narrow degree distribution.  Our analysis depends on having fixed in-degrees, and  we do not assume that networks are fully regular.  Both approaches lead to results that  hold approximately (for large enough $N$) when the in-degree is not fixed.

\paragraph{Average correlations between cells in the random network conditioned on first order connectivity}

As Fig.~\ref{F:rand_fig}B shows there is large variability around the mean excitatory-inhibitory cross-correlation function given 
by the leading order term of Eq.~\eqref{E:random_ct_av}.  Therefore, understanding the average cross-correlation between cells of given types does not necessarily provide much insight into the mechanisms that shape correlations on the level of individual cell pairs. Instead, we examine the average correlation  between a pair of cells  conditioned on their first order (direct) connectivity.  

We derive expressions for  first order conditional averages  correct to $\mathcal{O}(1/N^2)$ (See SI Sec. 2). The average cross-spectrum for a pair of  cells with indices $i \neq j$, conditioned on the value of the direct connections between them is
\begin{equation}\label{E:random_foc_av}
\begin{split}
\EVb{ \bfCt_{ij}^\infty| \bfJt_{ij}, \bfJt_{ij}}_{i \in X, j \in Y} &=  \Ct^0\left[ \underline{ \At \bfJt_{ij} + \At^*\bfJt_{ji}^* }+ \left(\frac{\At^2\mut}{1 - \At\mut}\right)\frac{\mut_Y}{N_Y}+ \left(\frac{\At^2\mut}{1 - \At\mut}\right)^*\frac{\mut_X^*}{N_X} \right.\\
&\hspace{1in}+ \left. \left|\frac{\At}{1-\At\mut} \right|^2 \mut_c \right] + \mathcal{O}(1/N^2).
\end{split}
\end{equation}
Here we set $\bfJt_{ij}=0$ if we condition on the absence of a connection $j \rightarrow i$, and 
 $\bfJt_{ij}=\Jt_{Y}/p$ if we condition on its presence.   The term $\bfJt_{ji}$ is set similarly.

Although Eq.~\eqref{E:random_foc_av} appears significantly more complicated than the cell-type averages given in Eq.~\eqref{E:random_ct_av}, they only differ in the underlined, first order terms.
The magnitude of  expected contributions from all higher order motifs  is unchanged and coincides with those in the all--to--all network.

Figure~\ref{F:rand_fig}C shows the mean cross-correlation function for  mutually coupled excitatory-inhibitory pairs.  Taking into account the mutual coupling significantly reduces variability (Compare with Fig.~\ref{F:rand_fig}B).  To quantify this reduction, we calculate the mean reduction in variability when correlation functions are computed conditioned on the connectivity between the cells. For a single network, the relative decrease in variability can be quantified using\footnote{We make use of the norm $||\cdot||_2$ defined by $||f||_2 = \left(\int |f|^2\right)^{1/2}$.}
$$
\mu_{\text{error}} = \frac{1}{N_T} \sum_{\substack{(i,j) \in T \\ i \, > \, j}} \frac{||\bfC_{ij}(\tau) - C_T^{\text{FOC}}(\tau)||_2}{||\bfC_{ij}(\tau) - C_T^{\text{CT}}(\tau) ||_2},
$$
where $T$ represents pairs of cells of a given type and connection (in the present example these are reciprocally coupled excitatory-inhibitory pairs), $N_T$ is the number of pairs of that type in the network, $C_T^{\text{CT}}(\tau)$ is the leading order  approximation of average correlations 
given only  the type of cells in $T$ (as in Eq.~\eqref{E:random_ct_av}), and $C_T^{\text{FOC}}(\tau)$
the leading order approximation to average correlations conditioned on the first order connectivity of class $T$ (as in Eq.~\eqref{E:random_foc_av}).  
Figure~\ref{F:rand_fig}D shows $\mu_{\text{error}}$  averaged over twenty networks. In particular, compare the reduction in variability when conditioning on bidirectional coupling between 
excitatory-inhibitory pairs shown in Figs.~\ref{F:rand_fig}B,C, with the corresponding relative error 
in Fig.~\ref{F:rand_fig}D (circled in red).


\section*{Discussion}

We have developed a general theoretical framework that can be used to 
describe the correlation structure in a network of spiking cells.  This theory
allows us to find tractable approximations of cross-correlation functions in terms of the network architecture and single cell 
response properties. The approach was originally used to study the population response
of the electrosensory lateral 
line lobe of weakly electric fish~\cite{Lindner:2005}. 
The key approximation relies on the assumption that the activity of cells in the network
can be represented by a mixed point and continuous stochastic process, as given in Eq.~\eqref{E:first}.  An iterative construction then leads to the expressions for approximate cross-correlations between pairs of cells given by Eq.~\eqref{E:freq_dom_lim}.

~\cite{Ostojic:2009} obtained formulas for cross-correlations that correspond to the first step in this
iterative construction,  given in Eq.~\eqref{E:first}.    Their approach captures  corrections due to direct coupling (first order terms) and direct common input (second order terms involving second powers of interaction kernels).  Our approach can 
be viewed as a generalization that also accounts for length two directed chains, along with all higher order corrections. As  Fig.~\ref{F:ee_fig} illustrates, these additional terms can yield significant contributions to the structure of cross-correlations. We note that another distinction between our approach and that of~\cite{Ostojic:2009} is that the present approach also allows us to calculate corrected auto-correlations, whereas the framework of~\cite{Ostojic:2009} did not allow for adjustments to auto-correlations.

\cite{Pernice:2011} analyzed the correlation structure in networks of interacting \emph{Hawkes processes}~\cite{Hawkes:1971-1,Hawkes:1971-2} using an approach similar to the one presented here.  In both cases, we represented correlations between cell pairs in terms of contributions 
of different connectivity motifs. However, our methods differ in important ways: while their expressions are exact for Hawkes processes, \cite{Pernice:2011} did not attempt to
match their results to more physiological cell models, and did not account for the response properties of individual cells, although  
it may be possible to approximately fit the Hawkes models to such models. 
\cite{Pernice:2011} examined only total spike count covariances, which equal the integrals of the cross-correlation functions. 
This leads to a loss of information about the temporal structure of correlations.
However, as they note, their approach can be extended to obtain complete cross-correlation functions for the Hawkes model.

To illustrate the power of the theory in analyzing the factors that shape correlations, we considered 
a number of simple examples for which the approximation 
Eq.~\eqref{E:freq_dom_lim} is tractable. We showed how 
 the theory can be used both to obtain intuition for the 
 effects that  shape correlations, and to quantify their impact. 
 In particular, we explained how only local connections affect correlations in a precisely tuned 
all--to--all network, and how strengthening inhibition may synchronize spiking activity.

Linear response methods are perturbative. For Eq.~\eqref{E:approx}
to be valid,  neurons need to respond approximately linearly to their inputs.  This will only 
be true if inputs are ``weak" relative to the dynamical operating point of the cell.  We 
assume the presence of a white noise background which linearizes the transfer function, although it is possible to extend the present methods to colored background noise~\cite{FourcaudTrocme:2003,Alijani:2011}.

 It may be surprising that an approach based on \emph{linear} response theory
can provide corrections  to cross-correlations of arbitrary order in network connectivity. 
Corrections to firing activity which are higher order in \emph{network connectivity}
emerge from the linear correction in Eq.~\eqref{E:vector_it}. A full expansion of firing activity would
include terms arising from corrections to the input-output
transfer of the individual cells beyond those captured by the linear response approximation.   Formally, including such terms
would be necessary to capture all contributions of a given order in network connectivity~\cite{Rangan:2009-1,Rangan:2009-2}. However,
the high accuracy demonstrated by our method indicates that, at least in some cases, these additional
terms are quite small. In short, our approximation neglects higher-order
corrections \emph{to the input-output transfer} of individual cells, which is compatible with the assumption
that the presence of background noise causes this transfer to be close to linear.


As expected from the preceding discussion, simulations suggest that, for IF neurons, our approximations become less accurate as cells receive progressively stronger inputs.  The ``physical" reasons for this loss of accuracy could be related to interactions between the "hard threshold" and incoming synaptic inputs with short timescales. While the theory will work for short synaptic timescales, it will improve for slower synaptic dynamics, limiting towards being essentially exact in the limit of arbitrarily long synaptic time constants (Note the improvement in the approximation for the FFI circuit for the slower timescale as exhibited in Fig.~\ref{F:ffi_fig}).  However, as we have shown, the theory remains applicable in a wide range of dynamical regimes, including relatively low noise, superthreshold regimes where cells exhibit strong oscillatory behavior. Moreover, the theory can yield accurate approximations of strong correlations due to coupling:  for the bidirectionally coupled excitatory circuit of Fig.~\ref{F:ee_fig}, the approximate cross-correlations match numerically obtained results even when correlation coefficients are large ($\bfrho_{E_1E_2}(\infty) \approx 0.8$ in the excitable regime, $\approx 0.5$ in the oscillatory regime).

Although we have demonstrated the theory using networks of integrate--and--fire neurons,
the approach is  widely applicable.  The linear response kernel and power spectrum 
for a general integrate and fire 
neuron model can be easily obtained~\cite{Richardson:2007}. In addition, it is also possible to obtain the rate, spectrum,
and rate response function for modulation of the mean conductance in the case of conductance-based (rather than current-based) synapses
(See \cite{Richardson:2009} and SI Sec. 3).  As the linear response kernel is directly related to
the spike triggered average~\cite{Gabbiani:1998, Ostojic:2009},  the proposed theoretical framework should be
applicable even to actual neurons whose responses are characterized experimentally.

The possibilities for future applications are numerous. For example, one open question is how well
the theory can predict correlations in the presence of adaptive currents~\cite{Richardson:2009}. In addition, the description of correlations in terms of architecture and response properties suggests the possibility of addressing the difficult inverse problem of inferring architectural properties from correlations~\cite{Nykamp:2007,Toyoizumi:2009,Paninski:2010,Ostojic:2009}. Overall, it is our hope that the present approach will prove a valuable tool in moving the computational neuroscience community towards a more complete understanding of the origin and impact of correlated activity in neuronal populations.

%
%
%


\section*{Methods}

\paragraph{Numerical methods}

Simulations were run in C++, and the stochastic differential equations were integrated with a standard Euler  method with a time-step of 0.01ms. General parameter values were as follows: $\tau_i = 20$ms, $E_{L,i} + E_i = -54$mV, $\sigma_i = \sqrt{12}$mV, $v_{th} = 20$mV, $v_r = -54$mV, $\tau_{ref} = 2$ms, $V_T = -52.5$mV, $\Delta_T = 1.4$mV, $\tau_E = 10$ms, $\tau_I = 5$ms, $\tau_{D,i} = 1$ms. Marginal statistics (firing rates, uncoupled power spectra and response functions) were obtained using the threshold integration method of~\cite{Richardson:2007} in MATLAB. All code is available upon request.

\paragraph{Calculation of stationary rates in a recurrent network}

The stationary firing rate of an IF neuron can be computed as a function of the mean and
intensity of internal noise ($E_i, \sigma_i$) and other cellular parameters ($\tau_i, E_{L_i}$, etc...)~\cite{Ricciardi:1979}. Denote the stationary firing rate
 of cell $i$ in the network by $r_i$, and by $r_{i,0}(E,\sigma)$ the stationary firing rate in
 the presence of white noise with mean $E$ and variance $\sigma^2$.  We keep the dependencies on other parameters are implicit.
The stationary rates, $r_i$, in the recurrent network without external input are determined
self-consistently by
$$
r_i = r_{i,0}(E_i',\sigma_i) =  r_{i,0}(E_i + \sum_j \bfW_{ij}r_j,\,\sigma_i) \quad i = 1,\ldots,N \; ,
$$
where we used  $\EVs{ f_i } = \sum_j \bfW_{ij} \EVs{ y_j } = \sum_j \bfW_{ij}r_j$. This equality holds because
the synaptic kernels, $\bfJ_{ij}$, were normalized to have area $\bfW_{ij}$.  These equations can typically
be solved by fixed-point iteration.

Note that this provides an effective mean input, $E_i'$, to each cell, but does not 
give adjustments to the variance, $\sigma_i$. We assume that the major impact of
recurrent input is reflected in $E_i'$, and ignore corrections to the cell response 
involving higher order statistics
of the input.  This approach is valid as long as  fluctuations in the recurrent input to each cell
are small compared to $\sigma_i$, and may break down otherwise~\cite{Brunel:1999}.

\paragraph{Derivation of Eq.~\eqref{E:first}:}

Equation~\eqref{E:first} can be obtained by a direct calculation:
\begin{equation*}
\begin{split}
&\EVs{ (y_i^1(t+\tau)-r_i)  (y_j^1(t)-r_j)} = \EVs{ (y_i^0(t+\tau)-r_i)(y_j^0(t)-r_j) } + \sum_k \EVs{(\bfK_{ik}*[y_k^0-r_k])(t+\tau)(y_j^0(t)-r_j)}\\
&\qquad\qquad+\sum_k\EVs{(y_i^0(t+\tau)-r_i)(\bfK_{jk}*[y_k^0-r_k])(t)} + \sum_{k,l}\EVs{(\bfK_{ik}*[y_k^0-r_k])(t+\tau)(\bfK_{jl}*[y_l^0 - r_l])(t)}\\
&\qquad=\delta_{ij}\bfC_{ii}^0(\tau) + (\bfK_{ij}*\bfC_{jj}^0)(\tau) + (\bfK_{ji}^-*\bfC_{ii}^0)(\tau) + \sum_k (\bfK_{ik}*\bfK_{jk}^-*\bfC_{kk}^0)(\tau)
\end{split}
\end{equation*}

\paragraph{Verification of Eq.~\eqref{E:matrix_rel_0}:}

First, Eq.~\eqref{E:vector_it} directly implies that
$$
\bfy^n(t) = \bfy^0(t) + \sum_{k=1}^n (\bfK^k *[\bfy^0 - \bfr])(t), \quad n \geq 0,
$$
which we may use to find, for each $n \geq 0$, 
\begin{equation} \label{E:matrix_rel_0_der}
\begin{split}
\bfC^{n}(\tau) &\equiv \EVs{ (\bfy^{n}(t+\tau)-\bfr)( \bfy^{n}(t)-\bfr )^T}\\
&= \EVs{(\bfy^0(t+\tau) - \bfr)(\bfy^0(t)-\bfr)^T} + \sum_{k=1}^n \EVs{(\bfK^k*[\bfy^0-\bfr])(t+\tau)(\bfy^0(t)-\bfr)^T}\\
&+ \sum_{k=1}^n \EVs{(\bfy^0(t+\tau)-\bfr)(\bfK^k*[\bfy^0-\bfr])^T(t)} + \sum_{k,l=1}^n \EVs{(\bfK^k*[\bfy^0-\bfr])(t+\tau)(\bfK^k*[\bfy^0-\bfr])^T(t)}\\
&= \bfC^0(\tau) + \sum_{k=1}^n ( \bfK^k * \bfC^0)(\tau) + \sum_{k=1}^n (\bfC^0 * (\bfK^-)^{kT})(\tau) + \sum_{k,l=1}^n ( \bfK^k*\bfC^0*(\bfK^-)^{lT})(\tau).
\end{split}
\end{equation}
Since $\bfK_{ij}^0(t) = \delta_{ij} \delta(t)$, Eq.~\eqref{E:matrix_rel_0_der} is equivalent to Eq.~\eqref{E:matrix_rel_0}.

\paragraph{Correction to statistics in the presence of an external white noise signals.}

Expression~\eqref{E:freq_dom_lim_ext} can be used to compute the statistics of the network response to inputs
$\eta_i(t)$ of finite variance.  As noted by~\cite{Lindner:2005}, when inputs have infinite variance additional corrections are necessary.  As a particular example, consider the case where the processes are
correlated white noise, i.e., when $\eta_i(t) = \sqrt{c}x_c(t) + \sqrt{1-c}x_i(t)$, where $x_c, x_i$ are  independent
white noise processes with variance $\sigma_e$. Then each $\eta_i$ is also a white noise process with intensity $\sigma_i^e$, but $\EVs{\eta_i(t+\tau)\eta_j(t)} = [\delta_{ij}\delta(\tau) + (1-\delta_{ij})c\delta(\tau)]\sigma_i^e$.
The firing rate of cell $i$ in response to this input is $r_i = r_0(E_i', \sqrt{(\sigma_i)^2 + (\sigma_i^e)^2})$, and the point
around which the response of the cell is linearized needs to be adjusted. 

Finally, we may apply an additional correction to the linear response approximation
of autocorrelations. For simplicity, we ignore coupling in Eq.~\eqref{E:freq_dom_lim_ext} (so that $\bfKt = 0$). Linear response predicts
that $\bfCt_{ii}(\omega) = \bfCt^0_{ii}(\omega;\sigma_i^2) + (\sigma_i^e)^2|\At_i(\omega)|^2$, where we have introduced explicit dependence on $\sigma_i^2$, the
variance of white noise being received by an IF neuron with power spectrum $\bfCt_{ii}^0(\omega ; \sigma_i^2)$, in the absence of the external signal. The approximation may be improved in this case by making the following substitution  in Eq.~\eqref{E:freq_dom_lim_ext}~\cite{Lindner:2005,Vilela:2009}:
$$
\bfCt^0_{ii}(\omega ; \sigma_i^2) + (\sigma_i^e)^2|\At_i(\omega)|^2 \quad\rightarrow\quad  \bfCt^0_{ii}(\omega ;  \sigma_i^2 + (\sigma_i^e)^2)
$$
The response function $A$ should be adjusted likewise.

\section*{Acknowledgements}

The authors thank Robert Rosenbaum, Brent Doiron and Srdjan Ostojic for many helpful discussions.
This work was supported by NSF grants DMS-0817649, DMS-1122094, and a Texas ARP/ATP award to KJ,  
as well as NSF Grants DMS-1122106, DMS-0818153, and a Burroughs Wellcome Career Award at the Scientific Interface to ESB.

\bibliography{paper1j,paper1e}

\newpage

\begin{table}
\centering
\begin{tabular}[h]{|l|p{5in}|}
\hline
$v_i, \tau_i, E_{L,i}, \sigma_i $ & Membrane potential, membrane time constant, leak reversal potential, and noise intensity of cell $i$.\smallskip\\

$E_i, \sigma_i$ & Mean and standard deviation of the background noise  for cell $i$. \smallskip\\

$v_{th},v_{r},\tau_{ref}$ & Membrane potential threshold, reset, and absolute refractory period for cells.\smallskip\\

$\psi(v),V_T,\Delta_T$ & Spike generating current, soft threshold and spike shape parameters for the IF model~\cite{FourcaudTrocme:2003}. \smallskip\\

$f_i(t), \eta_i(t)$ & Synaptic input from other cells in the network, and external input to cell $i$. \smallskip\\

$\tau_{S,i}, \tau_{D,i}$ & Synaptic time constant and delay for outputs of cell $i$. \smallskip\\

$y_i(t)$ & Spike train of cell $i$. \smallskip\\

$\bfW_{ij}$ & The $j \rightarrow i$ synaptic weight, describes the area under a single post-synaptic current for current-based synapses.\smallskip\\

$\bfJ_{ij}(t)$ & The $j\rightarrow i$ synaptic kernel - equals the synaptic weight $\bfW_{ij}$ multiplied times the synaptic filter for outputs of cell $j$.\smallskip\\

$\bfC_{ij}(\tau)$ & The cross-correlation function between cells $i,j$ defined by $C_{ij}(\tau) = \cov(y_i(t+\tau),y_j(t))$.\smallskip\\

$N_{y_i}(t,t+T),\bfrho_{ij}(T)$ & Spike count for cell $i$, and spike count correlation coefficient for cells $i,j$ over windows of length $T$.\smallskip\\

$r_i, A_i(t), \bfC^0_{ii}$ & Stationary rate, linear response kernel and uncoupled auto-correlation function for cell $i$j.\smallskip\\

$\bfK_{ij}(t)$ & The $j\rightarrow i$ interaction kernel - describes how the firing activity of cell $i$ is perturbed by an input spike from cell $j$. It is defined by $\bfK_{ij}(t) = (A_i * \bfJ_{ij})(t)$.\smallskip\\

$\bfy^n_i(t), \bfC^n_{ij}(t)$ & The $n^{th}$ order approximation of the activity of cell $i$ in a network which accounts for directed paths through the network graph up to length $n$ ending at cell $i$, and the cross-correlation between the $n^{th}$ order approximations of the activity of cells $i,j$.\smallskip\\

$g(t), \tilde{g}(\omega)$ & $\tilde{g}(\omega)$ is the Fourier transform of $g(t)$ with the convention 
$$\tilde{g}(\omega) = \mathcal{F}[g](\omega) \equiv \int_{-\infty}^\infty e^{-2 \pi i \omega t}g(t)dt$$ \smallskip\\

\hline
\end{tabular}
\caption{{\bf Notation used in the text.}}
\label{T:notation}
\end{table}

\end{document}